\documentclass[a4paper,fleqn,usenatbib]{mnras}

\usepackage{newtxtext,newtxmath}
\usepackage[flushleft]{threeparttable}
\usepackage{times,epsfig,natbib,graphics,longtable,lscape}
\usepackage[edges]{forest}
\usepackage{pdflscape}
\usepackage{subfig}
\usepackage[export]{adjustbox}
\usepackage[T1]{fontenc}
\usepackage{ae,aecompl}

\usepackage{nccmath}
\usepackage{multicol}

\title[Optical/$\gamma$-ray blazar flare correlations]{Optical/$\gamma$-ray blazar flare correlations: understanding the high-energy emission process using ASAS-SN and Fermi light curves}

\author[de Jaeger et al.]
{T. de Jaeger$^{1}$\thanks{E-mail: dejaeger@hawaii.edu},
B. J. Shappee$^{1}$,
C. S. Kochanek$^{2,3}$,
J.~T. Hinkle$^{1}$,
S. Garrappa$^{4}$,
\newauthor
I. Liodakis$^{5}$,
A. Franckowiak$^{4}$, 
K. Z. Stanek$^{2,3}$,
J.~F.~Beacom$^{6,2,3}$,
J.~L.~Prieto$^{7,8}$\\
$^{1}$ Institute for Astronomy, University of Hawaii, 2680 Woodlawn Drive, Honolulu, HI 96822, USA.\\
$^{2}$ Department of Astronomy, The Ohio State University, 140 W. 18th Avenue, Columbus, OH 43210, USA.\\
$^{3}$ Center for Cosmology and AstroParticle Physics (CCAPP), The Ohio State University, 191 W. Woodruff Avenue, Columbus, OH 43210, USA.\\
$^{4}$ Fakult{\"a}t f{\"u}r Physik \& Astronomie, Ruhr-Universit{\"a}t Bochum, D-44780 Bochum, Germany.\\
$^{5}$ Finnish Centre for Astronomy with ESO, 20014 University of Turku, Finland.\\
$^{6}$ Department of Physics, The Ohio State University, 191 W. Woodruff Ave., Columbus, OH 43210, USA.\\
$^{7}$ N\'ucleo de Astronom\'ia de la Facultad de Ingenier\'ia y Ciencias, Universidad Diego Portales, Av. Ej \'ercito 441, Santiago, Chile.\\
$^{8}$ Millennium Institute of Astrophysics, Nuncio Monsenor S\'otero Sanz 100, Providencia 8320000, Santiago, Chile.}

\date{}

\pubyear{2022}

\begin{document}
\label{firstpage}
\pagerange{\pageref{firstpage}--\pageref{lastpage}}

\maketitle

\begin{abstract}
\noindent
Using blazar light curves from the optical All-Sky Automated Survey for Supernovae (ASAS-SN) and the $\gamma$-ray \textit{Fermi}-LAT telescope, we performed the most extensive statistical correlation study between both bands, using a sample of 1,180 blazars. This is almost an order of magnitude larger than other recent studies. Blazars represent more than 98\% of the AGNs detected by \textit{Fermi}-LAT and are the brightest $\gamma$-ray sources in the extragalactic sky. They are essential for studying the physical properties of astrophysical jets from central black holes. However, their $\gamma$-ray flare mechanism is not fully understood. Multi-wavelength correlations help constrain the dominant mechanisms of blazar variability and the emission source region. For this purpose, we search for temporal relationships between optical and $\gamma$-ray bands. Using a Bayesian Block Decomposition, we detect 1414 optical and 510 $\gamma$-ray flares, we find a strong correlation between both bands. Among all the flares, we find 321 correlated flares from 133 blazars, and derive an average rest-frame time delay of only 1.1$_{-8.5}^{+7.1}$ days, with no difference between the flat-spectrum radio quasars, BL Lacertae-like objects or low, intermediate, and high-synchrotron peaked blazar classes. Our time-delay limit supports leptonic single-zone model as the driver for non-orphan flares. Limiting our search to well-defined light curves and removing 976 potential but unclear ``orphan'' flares, we find 191 (13\%) and 115 (22\%) clear ``orphan'' optical and $\gamma$-ray flares. The presence of ``orphan'' flares in both bands challenges the standard one-zone blazar flare leptonic model and suggests multi-zone synchrotron sites or a hadronic model for some blazars.

\end{abstract}



\section{Introduction}
Blazars are a subclass of radio-loud active galactic nuclei (AGNs) characterised by a jet pointing within a few degrees of the observer's line of sight \citep{blandford78,antonucci93,urry95}. Due to the nearly-aligned viewing angle, we observe strong relativistic effects such as beaming of the emitted power, causing blazars to be the brightest $\gamma$-ray sources in the extragalactic sky. They represent more than 98\% of the AGNs detected by the Large Area Telescope (LAT) onboard the \textit{Fermi} $\gamma$-ray space observatory \citep{atwood09,acero15,abdollahi20} and are ideal objects for studying the poorly understood physics of astrophysical jets.

Based on the strength of their optical emission lines, blazars are subdivided into flat-spectrum radio quasars (FSRQs) and BL Lacertae (BL Lac) objects \citep{urry95}. FSRQs have prominent broad emission lines, while BL Lacs have blue nearly featureless optical spectra with emission line equivalent width $<$ 5\AA \citep{stickel91}. Recently, \citet{abdo10} and \citet{ghisellini11} proposed an alternative classification based on the peak of their spectral energy distribution (SED): low-synchrotron peaked blazars (LSP; $\nu_{\rm peak}$ $\leq$ 10$^{14}$ Hz), intermediate synchrotron peaked blazars (ISP; 10$^{14}$ $\leq$ $\nu_{\rm peak}$ $\leq$ 10$^{15}$ Hz), and high-synchrotron peaked blazars (HSP; $\nu_{\rm peak}$ $>$ 10$^{15}$ Hz). The HSP and ISP blazar groups are mostly BL Lacs, while the LSP class is dominated by FSRQs.

All blazars emit across the electromagnetic spectrum. Their SED consists of two broad non-thermal peaks, one at radio to ultraviolet wavelengths and the second at X-ray to $\gamma$-ray energies \citep{impey88,fossati98,marscher08}. The low-frequency peak is generally agreed to be synchrotron emission from relativistic electrons spiralling in the jet magnetic field \citep{urry82,impey88}. However, the origin of the higher-energy peak is not fully understood. It can be explained by leptonic, hadronic \citep{mucke03,bottcher13,madejski16}, or hybrid models (see \citealt{bottcher19} for a review). 

In the leptonic model, electron pairs dominate the high-energy emissions, as the protons within the outflow are not accelerated to sufficiently high energies to make a significant contribution. Therefore, GeV-TeV $\gamma$-rays are due to inverse Compton scattering (ICS) of low-energy (IR-optical-UV) target photons by the same ultra-relativistic electrons producing the synchrotron emission at lower frequencies \citep{maraschi92,sikora94}. Since the low and high-energy peaks are produced by the same population of electrons, the leptonic model predicts strong correlations between the optical and $\gamma$-ray emission. The origin of the seed photons is not clear. They could be produced within the jet (synchrotron self-Compton, SSC; e.g., \citealt{marscher85,ghisellini89,maraschi92,bloom96,chiang02,arbeiter05}) or from external sources (external Compton, EC) such as the accretion disk, the ``dusty torus'', or the broad-line region \citep{dermer93,sikora94,ghisellini96,boettcher97,blazejowski00,dermer09}. 

The absence of thermal emission in BL Lac blazar flares favours a one-zone leptonic SSC model, where a single emitting region with constant and homogeneous parameters dominates the emission \citep{costamante02}. One-zone leptonic SSC models are interesting because of their simplicity and their limited number of free parameters. However, unlike BL Lac, the FSRQ SED is not easily described by a simple one-zone SSC model. It is best modelled with a sum of the contribution from the SSC and the scattering of external photons with the EC contribution dominant during strong $\gamma$-ray flares \citep{sikora94,ghisellini12,williamson14,bottcher19}.

In hadronic models, the high-energy emission is associated with ultra-relativistic protons emitted by proton-synchrotron radiation \citep{aharonian00,mucke01,mucke03,liodakis20} or photo-pion production \citep{mannheim93}. While strong long-term correlated variability between the low-energy bands and $\gamma$-rays is generally not expected for the hadronic model, we expect a strong correlation between the $\gamma$-ray and the neutrino fluxes, as the pion decay leads to the production of ultra-high-energy photons and neutrinos \citep{icecube2018}. Note also, there could still be some correlation between optical and $\gamma$-ray emission, because the hadronic processes that produce high-energy $\gamma$-rays also produce high-energy electrons, which act just like those in leptonic models. However, the corresponding low-energy emission can be overwhelmed by other sources.

Understanding the dominant process for blazar's $\gamma$-ray emissions and the physics of the astrophysical jets can be done through two complementary approaches:

(i) The first is broadband SED modeling of individual objects such as 3C 454.3, Mrk 421, Mrk 501 or PKS 2155-304. Unfortunately, leptonic, hadronic, and lepto-hadronic scenarios have all been able to successfully reproduce the SEDs of blazars (e.g., \citealt{punch92,dermer93,mannheim93,urry97,ghisellini98,mucke01,blazejowski05,aharonian07,albert07,jorstad10,bottcher13,cerruti19}), so it has been difficult to constrain the dominant process. Obtaining the SED also requires intensive multi-wavelength observational campaigns for each source. 

(ii) An alternative method is to study flux variations and search for temporal correlations between the two major emission components (optical and $\gamma$-ray). This method can now be applied to a large number of blazars thanks to the well-sampled $\gamma$-ray light curves from \textit{Fermi}-LAT and the optical light curves from new generations of optical surveys like the All-Sky Automated Survey for Supernovae (ASAS-SN; \citealt{Shappee14,kochanek17b}). This allows a significant expansion in the number of optically monitored blazars compared to previous campaigns studying individual objects such as Yale/SMARTS \citep{chatterjee12,bonning12} or the Katzman Automatic Imaging Telescope (KAIT; \citealt{filippenko01,cohen14,liodakis19}).

Recent multi-wavelength correlation investigations of large blazar samples have constrained the dominant mechanisms driving blazar variability. For example, \citet{liodakis19} studied 178 \textit{Fermi}-LAT blazars regularly monitored by KAIT and SMARTS and found strong optical/$\gamma$-ray correlations with time delays of only a few days (1 to 30 days), supporting the leptonic models and confirming previous studies with smaller samples \citep{bonning12,chatterjee12,hovatta14,cohen14,liodakis18}. However, the presence of $\gamma$-ray flares without a low-energy counterpart and optical flares without a high-energy counterpart (known as ``orphan'' flares) in the same blazar sample challenges the standard one-zone leptonic blazar flare model \citep{chatterjee12,bottcher13,cohen14,liodakis19}. There are several models  for the origin of orphan flares (see \citealt{bottcher19} for a review), including a hadronic synchrotron mirror model in which a flare is due to the interaction of relativistic protons within the jet with external photons from the reflected electron-synchrotron emission from a nearby external obstacle \citep{bottcher05}, a two-zone model with a site of $\gamma$-ray production situated between the broad-line region and the jet recollimation shock \citep{macdonald15}, an encounter between relativistic blobs in the jet and a luminous star \citep{banasinski16}, the effects of the magnetic fields \citep{joshi16,sobacchi21}, or a stochastic dissipation model \citep{wang22}.

In this work, we investigate the question of the high-energy emission process by looking for optical/$\gamma$-ray flare correlations for 1,180 blazar light curves from the \textit{Fermi}-LAT Collaboration database \citep{abdollahi22}. This sample represents the most extensive and homogeneous statistical study of blazar flares. Our paper is organised as follows. Section 2 describes our optical and $\gamma$-ray data. Section 3 presents our methodology to search for optical-$\gamma$-ray flares and measure rest-frame time lag. Section 4 presents our our results. Finally, Section 5 contains a summary and conclusions. Appendix A and B present the sample and ``orphan'' flare information.

\section{Data Samples}\label{sec:sample}

We selected all the sources from the 12-year \textit{Fermi}-LAT point source (4FGL-DR3) catalogue \citep{abdollahi22} with a variability index greater than 24.725. For these sources, the chance of being a steady source is $<$ 1\%. This leads to a sample of 1,695 sources ($\sim$ 25\% of the 4FGL-DR3 6,659 source catalogue). Then, we remove all sources with any non-zero entry in the analysis flags column indicating
 that they are affected by systematic errors \citep{abdollahi22}. Finally, we select only the blazars and blazar candidates --- the sources classified as FSRQ, BL Lac, or blazar candidates of uncertain type (BCU) --- with light curves in the Fermi LAT Light Curve Repository (LCR\footnote{\url{https://fermi.gsfc.nasa.gov/ssc/data/access/lat/LightCurveRepository/about.html.}}). This leads to a final sample of 1,180 sources consisting of 503 FSRQs, 437 BL Lacs, and 240 BCU (814 LSP, 123 ISP, and 127 HSP) sources. This source list is provided in Table \ref{tab:sample}. 

\subsection{$\gamma$-ray light curves}

The $\gamma$-ray light curves were obtained from the LAT Light Curve Repository website and included data binned at three day, one-week, and one-month intervals. For our analysis, we use the three-day binning to detect short flares. A full description of the data reduction is found on the \textit{Fermi}-LAT Light Curve Repository website and in \citet{abdollahi22}. Therefore, we present only a brief description here.

The LAT sources are characterised using an unbinned maximum likelihood analysis \citep{abdo09} in which the complete spatial and spectral information of each photon is used in the maximum likelihood optimisation. Light curves from the LCR were created using the standard Fermi Tools 1.4.7 analysis suite with the $P8R3\_SOURCE\_V2$ instrument response functions on $P8R3\_SOURCE$ class photons selected over an energy range spanning 100 MeV--100 GeV. To select the photons, a region of interest with a radius of 12$^{\circ}$  is centred on the source localisation with a zenith angle cut of 90$^{\circ}$ to prevent contamination from $\gamma$-rays from the Earth limb. Then, the region of interest is fitted as a point source plus a model including the diffuse $\gamma$-ray emission from our Galaxy (template: $gll\_iem\_v07$), the isotropic template to account for all remaining isotropic emission (template: $iso\_P8R3\_SOURCE\_V2\_v1.txt$), the Sun and Moon steady emission templates, and all point-like and extended sources from DR3. The significance of the source detections are quantified by a test statistic (TS). Here, we select a detection criterion such that the maximum-likelihood TS exceeds four ($\sim$2$\sigma$). Note that no assumptions about the spectral shape of the gamma-ray sources is made and the photon index fits are fixed.

\subsection{Optical light curves}

ASAS-SN is a ground-based survey able to observe the entire visible sky daily to a depth of $g = 18.5$ mag \citep{Shappee14,kochanek17b}. Starting in 2013, with its first unit (Brutus) located on Haleakal\=a in Hawaii (USA), ASAS-SN is now composed of five stations in both hemispheres. Two units at the Cerro Tololo International Observatory in Chile (Cassius and Paczynski), one at McDonald Observatory in Texas, USA (Leavitt), and finally, one at the South African Astrophysical Observatory in Sutherland, South Africa (Payne-Gaposchkin). Each unit consists of four 14-cm aperture Nikon telephoto lenses with back-illuminated, 2048$^{2}$ CCD cameras having a 4.47 by 4.47-degree field of view. Until late 2018, Brutus and Cassius units used a $V$-band filter and were switched to $g$-band filters after roughly one year of $V$- and $g$-band overlap with the three new units. 

For each ASAS-SN field, we take three dithered 90-second exposures. All the images are processed by the \textit{ISIS} \citep{alard98,alard00} image substraction ASAS-SN pipeline (\citealt{Shappee14}; \citealt{kochanek17b}; \citealt{dejaeger22a}; Hart et al. in prep). For photometry, we use the subtracted image with the reference flux of the source added back to the light curve (see \url{https://www.astronomy.ohio-state.edu/asassn/public/examples.shtml}). With a 2-pixel radius ($\sim$ 16\arcsec) and the \textit{IRAF apphot} package, we conduct aperture photometry on the coadded image-subtracted data for each nightly epoch. Note that the blazar flux in the reference image leads to small offsets in the light curve for each ASAS-SN camera. To correct the light curves, we derive an average offset by comparing the average flux of each camera during the overlapping period. As the ``reference'' flux, we choose the camera with a longer time coverage range. Then, we apply the different cameras' offsets to the rest of the light curve. Finally, we bin all the light curves to a three day cadence to be consistent with the $\gamma$-ray light curves.

\begin{figure}
\centering
\includegraphics[width=0.5\textwidth]{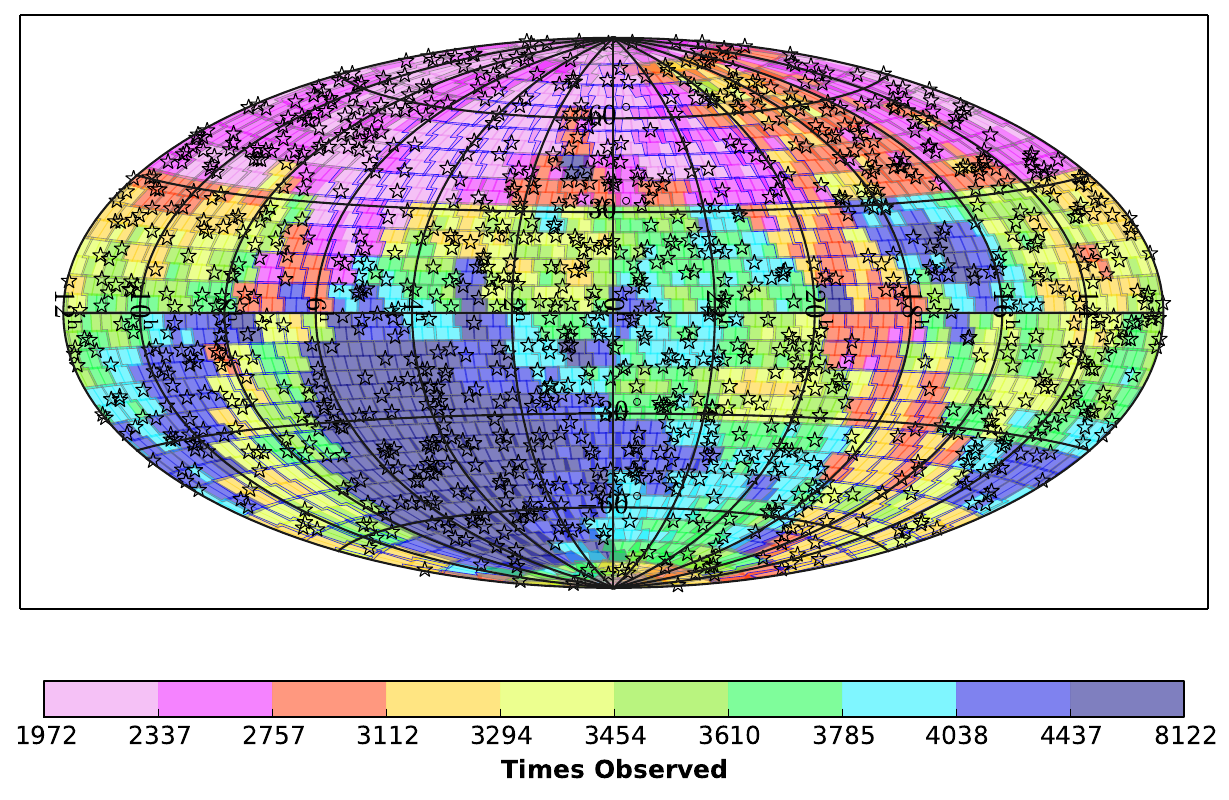}
\caption{The all-sky, high-cadence, decade-long coverage of both ASAS-SN and \textit{Fermi}-LAT makes the pairing of these two observational datasets compelling. ASAS-SN covers the entire visible sky with at least 1,970 epochs and more than 8,100 for some number of sky regions (as of October 11, 2022). The colour bar indicates the number of ASAS-SN epochs and the black stars indicate the 1,180 \textit{Fermi}-LAT sources.}
\label{fig:ASASSN_coverage}%
\end{figure}

ASAS-SN is ideal for searching for blazar flare correlations as it has two unique advantages over other optical surveys:\\
i) Its observational baseline begins in 2013, covering the entire sky. For comparison, ZTF has observed only the Northern hemisphere since 2018 \citep{bellm19}. As seen in Figure \ref{fig:ASASSN_coverage}, there are at least 1,970 and up to 8,100 observations\footnote{One epoch consists of three observations of 90 seconds.} of every point in the sky.\\
ii) Unlike targeted optical blazar monitoring programs, it has observations of all the \textit{Fermi}-LAT sources.

\noindent Those two unique features allow us to do the most extensive statistical study of blazar flares to date with minimal selection biased, as we observe all the blazars independently of their properties or localisation.

\section{Search for optical/$\gamma$-ray correlations}

In this section, we detail our methodology to search for flares and how we derive the time lag between the optical and the $\gamma$-ray emissions.

\subsection{Flares}\label{txt:flares}

In blazar studies, one of the most important goals is to search for and characterise flares. Flares are prominent outbursts lasting for days to months surrounded by period of relative quiescence. Unfortunately, there is no a consensus on a quantitative definition of a flare. Here, we give a non-exhaustive list of the approaches used in the literature:

\begin{enumerate}

\item{A flare is simply defined as the period when the flux level exceeds <F>+3$\sigma_{w}$ where <F> is the average flux and $\sigma_{w}$ is the weighted standard deviation \citep{williamson14}.}

\item{A flare is a contiguous period of time associated with a given flux peak, during which the flux exceeds half of the peak value, and this lower limit is attained exactly twice at the beginning and the end \citep{nalewajko13}.}

\item{The blazar light curves are decomposed into individual flares \citep{valtaoja99}, each described by an exponential rise and decay \citep{chatterjee08,jorstad10,chatterjee12,liodakis18b,roy19}.}

\item{A flare is defined using the Bayesian block algorithm \citep{scargle98,scargle13}. First, the Bayesian block algorithm is used to segment the blazar light curve into blocks with statistically constant fluxes. Then, we choose a flux threshold above which a block is designated as a flare \citep{liodakis18,meyer19,adams22}.}

\item{A flare is the period of time where the fractional variability amplitude (Fvar), which characterises the flux variability properties of a blazar, is above a threshold \citep{vaughan03}.}

\end{enumerate}

For this study, we use Bayesian Block Decomposition (BBD; \citealt{scargle13,scargle98}), as it seems to be the most objective way to identify strong flares, for both the $\gamma$-ray and optical light curves. Additionally, it can identify significant data series changes independently of gaps or exposure variations and does not need a prior for the flare timescale. BBD determines an optimal binning for the data, where the bin sizes do not have the same width but each bin has no statistically significant flux variations. This method requires a false-positive rate (p$_{0}$) associated with the prior estimate of the number of bins, ncrprior, defined as $4-\mathrm{log}(73.53p_{0}N^{-0.478})$, where N is the number of photometric points. We choose a false-positive rate for the optical and the $\gamma$-ray data of p$_{0}$=0.01.

After running the BBD, the non-flaring and flaring levels are identified using a three-step procedure following \citet[see also \citealt{wagner22} for \textit{Python} codes]{meyer19}.

i) We define the flares (i.e., the average flux in the block; $\overline{F_{blocks}}$) as all the local maxima\footnote{All the blocks that are higher than both the previous and subsequent blocks.} with a flux level higher than $\widetilde{F}$+$\sigma_{F}$ where $\widetilde{F}$ is the median flux and $\sigma_{F}$ is the standard deviation of the whole light curve. 

ii) We regroup blocks using the HOP algorithm \citep{eisenstein98} based on a bottom-up hill-climbing concept. We proceed downward from the peak, and every block subsequently lower in both directions (left and right) belongs to that peak. The flux exceeding our quiescent level ($\widetilde{F}$ $+$ $\sigma_{F}$) determines the start/end of the flare. Note that using the median flux for our quiescent level yields an upward bias for the true quiescent level estimate. However, this has little effect for our results, as we focus only on the most prominent flares (see Section \ref{txt:tlag}). 

iii) For our work, we study only the prominent flares by only selecting the flares with at least two flux points ($N>2$) where
\begin{ceqn}
\begin{align}
\overline{F_{blocks}} > 3\times \frac{\overline{F_{blocks}}-\widetilde{F}}{\sigma_{F}} \times \sqrt{N}.
\label{eq:flares}
\end{align}
\end{ceqn}

\noindent Finally, we visually inspect all the flare candidates and remove the bad candidates: flares with large photometric errors, optical flares due to seasonal gap edges (e.g., higher airmass), or optical flares seen only in one optical band ($V$ or $g$) during the $V/g$ band overlap period. Then, we classify each source  as: 

(i) Having at least one flare in both bands (both).

(ii) Only in optical (opt).

(iii) Only in $\gamma$-ray (gam).

(iv) Without flares (none).\\

\begin{figure*}
\centering
\includegraphics[width=1.0\textwidth]{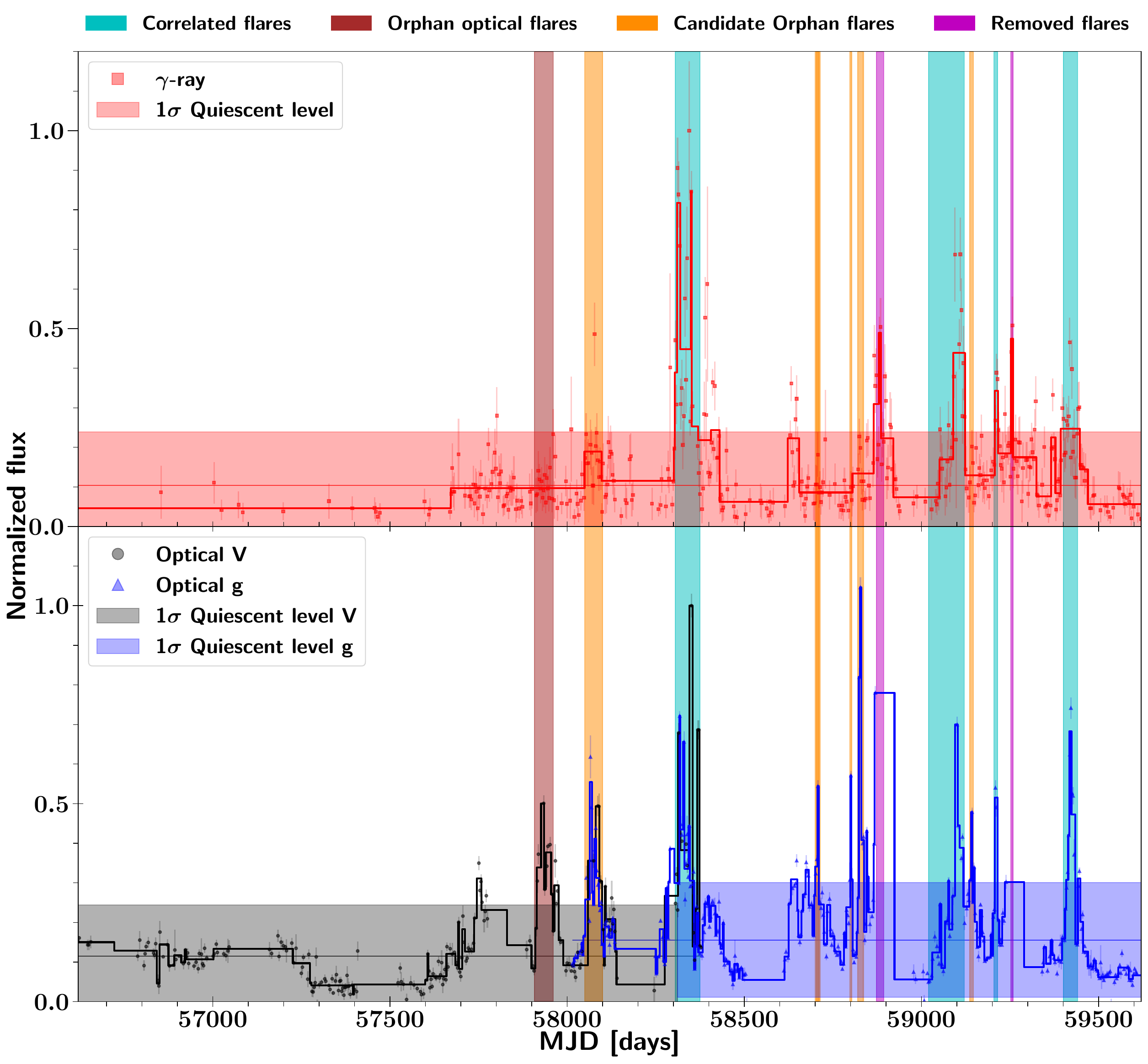}
\caption{\textit{Upper panel}: \textit{Fermi}-LAT $\gamma$-ray light curves for J0038.2-2459 and its Bayesian Block Decomposition (solid line), together with the identified HOP groups shown by the shaded regions. \textit{Bottom panel}: Optical $V$ and $g$ light curves from ASAS-SN and their BBD are shown respectively in red and black. In both panels, the vertical cyan, brown, orange and magenta-shaded regions represent the correlated flares, the flares seen the in optical but not in $\gamma$-rays (``orphan'' optical flares), the ``orphan'' candidate flares, and the flares removed after visual inspection, respectively.}
\label{fig:BBD_HOP}
\end{figure*}

\noindent Table \ref{tab:sample} shows a summary of the results of our visual inspections. Figure \ref{fig:BBD_HOP} shows an example of a blazar with flares detected in both bands after applying our methodology. We highlight correlated flares, ``orphan'' flares, candidate ``orphan'' flares, and the flares removed after our visual inspection in different colours. We select only the clearest examples for our ``orphan'' flare analysis, as explained in Section \ref{txt:res_orph}. For example, we do not consider the first ``orphan'' flare candidate at $\sim$ MJD 58050 as a true ``orphan'' flare. At the same epoch, the $\gamma$-ray light curve presents a small flux rise, not enough to be considered a flare according to our definition but enough to consider the optical flare as not an ``orphan''. Note also that the two ``orphan'' $\gamma$-ray flares detected by our methodology (magenta) are removed from the sample because they occurred during an optical observational gap.

\subsection{Time lags}\label{txt:intro_tlag}

To resolve the structure of relativistic jets and the localisation of the emission along the length of the jet, we can study the time lags between flares in different energy bands. For example, in the leptonic single-zone model (SSC and EC), one expects time delays of only a few days because the same electrons produce flares in both bands. For example, we find short delays when the EC is the dominant mechanism due to the stratification with the distance from the black hole in the radiation field and the profile of the magnetic field \citep{janiak12}. However, for the hadronic models, the low- and high-energy SED peaks vary independently and we do not expect a correlation between the optical and gamma-ray flares \citep{sikora94,sokolov04,cohen14,liodakis19,bottcher19}. Different works using different techniques and samples have found a strong temporal correlation between optical and $\gamma$-ray flares supporting leptonic models for the blazar flares. In particular, \citet{bonning12}, \citet{cohen14}, \citet{liodakis19}, and \citet{bhatta21} only found short time delays of the order of days to tens of days.

In the literature, two methods are commonly used to study the cross-correlation between time series and estimate lags:

(i) Discrete Correlation Function (DCF): this method is based on the discrete correlation function \citep{edelson88}, which, unlike the auto-correlation function, allows an estimates from unevenly sampled data.

(ii) Z-transformed discrete correlation function (ZDCF): this is a modified version of the DCF \citep{alexander97,alexander13} where the cross-correlation coefficients, r, are z-transformed using Fisher transformations. Unlike the DCF, the data are rebinned with an equal numbers of data points in each lag bin. The ZDCF reduces biases and gives better estimates of the uncertainties.

In general, one computes the DCF and the ZDCF for a range of time lags. Then, the correlation peak is fit with a Gaussian to determine the time lag and its uncertainty. 

\begin{figure*}
\centering
\includegraphics[width=\textwidth]{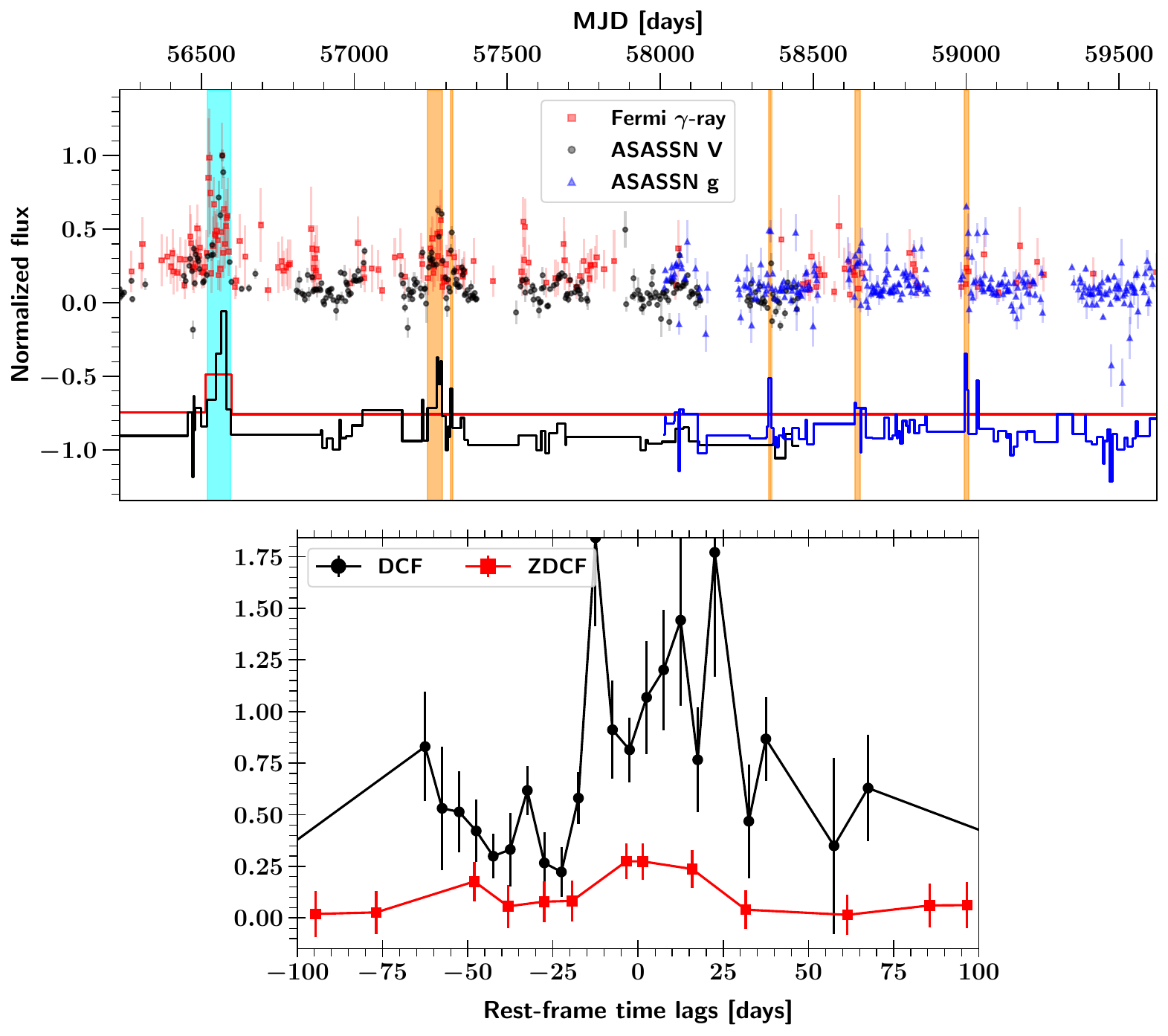}
\caption{\textit{Upper panel}: Correlated flares in the normalised light curves of J0050.4-0452 between the \textit{Fermi}-LAT $\gamma$-ray (red squares) and the optical $V$-band (black dots), and $g$-band (blue triangles). The cyan shaded region indicates the flares in both bands and used to derive the time lag. The orange shaded regions represent the flare detected in the optical light curves but not in the $\gamma$-ray light curve. The black, red and blue lines are the Bayesian Block Decompositions for each band shifted downwards for clarity. \textit{Bottom panel}: Discrete correlation function (DCF; in black) and Z-transformed discrete correlation function (ZDCF; in red) between the optical and $\gamma$-ray light curves. A positive observed time lag corresponds to the $\gamma$-ray emission leading to the optical emission.}
\label{fig:dcf}

\end{figure*}

Here, we focus on the lags for individual correlated flares instead of cross correlations the entire light curve. By looking only at flares, we are less sensitive to seasonal systematics, especially for sources near the detection limits of ASAS-SN and Fermi. This technique is better for objects where the DCF or the ZDCF curves do not show a peak in the correlation function, but the optical and $\gamma$-ray light curves clearly have correlated flares. An example is shown in Figure \ref{fig:dcf}. For this source, both DCF and ZDCF correlation functions lack a well-defined peak. However, both light curves clearly have correlated flares. For example, around MJD 56550, a flare is visible in both bands, with the $\gamma$-ray emission leading the optical. Another possible less significant correlated flare is visible around MJD 57250 but it was not detected by the Bayesian Block Decomposition of the $\gamma$-ray light curve. 

For all the blazars, we define the rest-frame time lag ($\tau_{lag}$) as the difference in the epochs of the Bayesian block peaks (the middle of the block) and the associated uncertainty as half the block widths. For the object presented in Figure \ref{fig:dcf}, the $\gamma$-ray and optical observed flare epochs are respectively MJD 56557.25 $\pm$ 42.8 and MJD 56574.12 $\pm$ 38.2. So the rest-frame is $\tau_{lag}$= 8.8 $\pm$ 21.1 days. A positive $\tau_{lag}$ corresponds to the $\gamma$-ray emission leading to the optical emission.

\section{Results}\label{txt:RESULTS}

In this section, we investigate if flaring activities are stronger in one of the bands. One metric for this is the number of flares (correlated and ``orphan'') observed in each band and for each blazar class. Then, we measure the time lag between optical and $\gamma$-ray emissions.

\subsection{Flares}\label{txt:res_flares}

After applying our search methodology from Section \ref{txt:flares}, we end up with a total of 2584 flares in 421 blazars\footnote{Blazars labeled as ``opt'', ``both'' or ``gam'' in Table \ref{tab:sample}.}: 1624 optical and 960 $\gamma$-ray flares. If we remove all the flares in observational gaps (at least 50 days without data) to keep only flares with overlapping optical and $\gamma$-ray observation periods---the total decreases to 1924 flares: 1414 optical and 510 $\gamma$-ray flares. Roughly $\sim$ 50\% of $\gamma$-ray flares occur during optical observing gaps. This is consistent with the 43\% found by \citet{liodakis19}. 

\begin{figure*}
\centering
\subfloat{\includegraphics[width=0.45\textwidth]{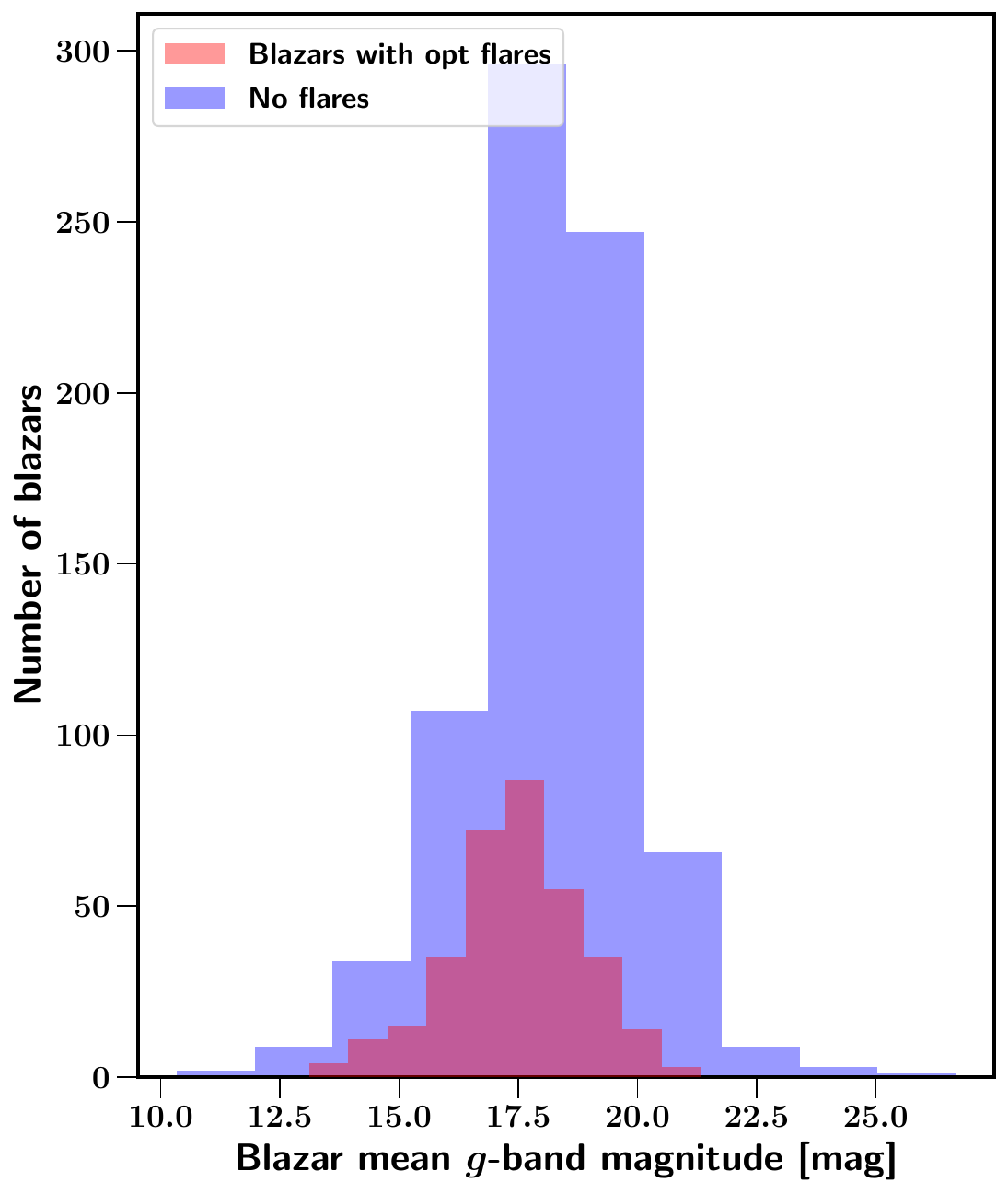}}
\subfloat{\includegraphics[width=0.45\textwidth]{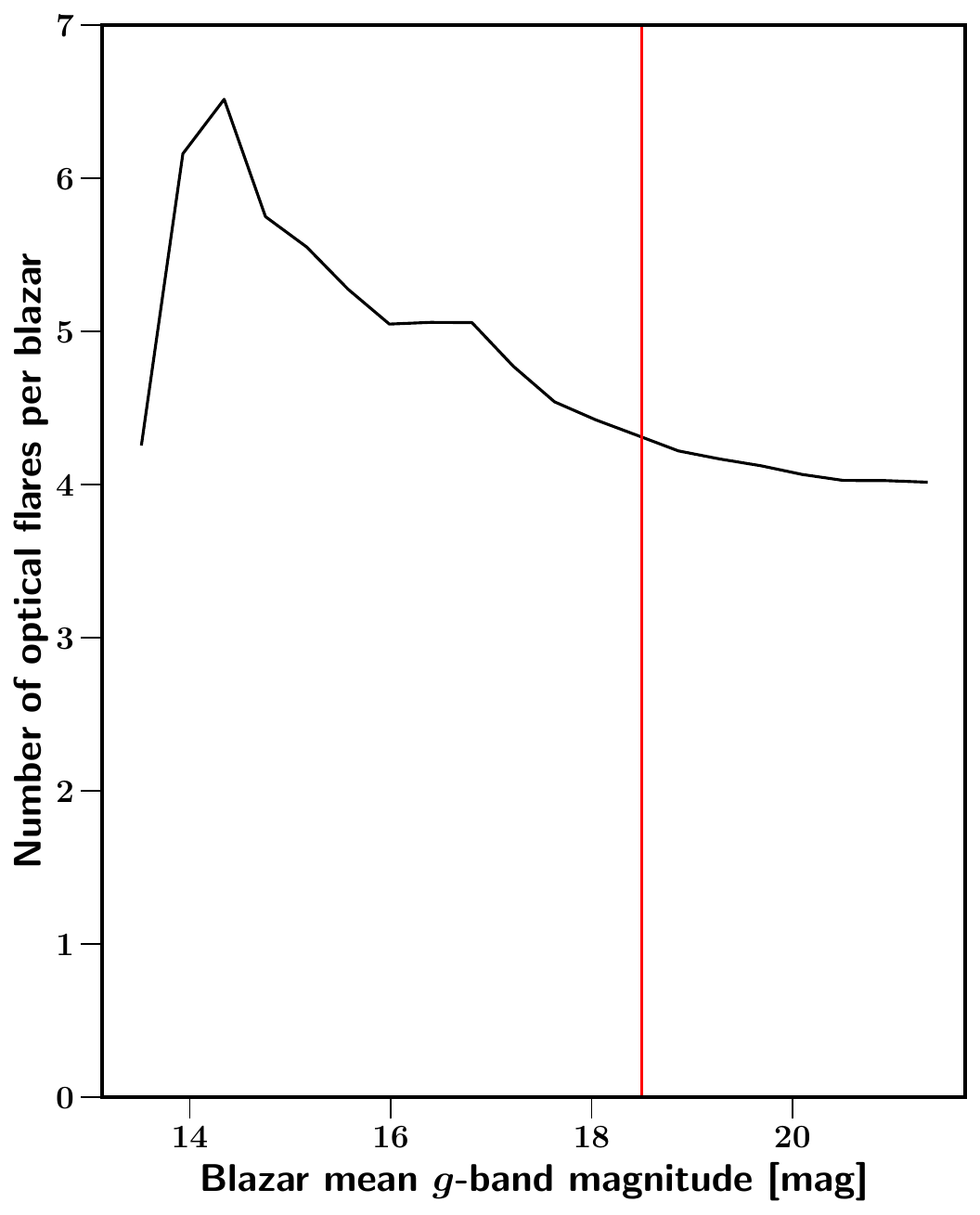}}
\caption{Optical flaring versus optical magnitude. \textit{left:} Histogram of the mean $g$-band blazar magnitude for the objects with at least one optical flare (red) and the sources without flares (blue). \textit{left:} Number of optical flares versus the mean $g$-band blazar magnitude. The red vertical line represents our magnitude cut at $g=18.5$ mag.}
    \label{fig:mag_lim}
\end{figure*}

\begin{figure*}
\centering
\includegraphics[width=1.0\textwidth]{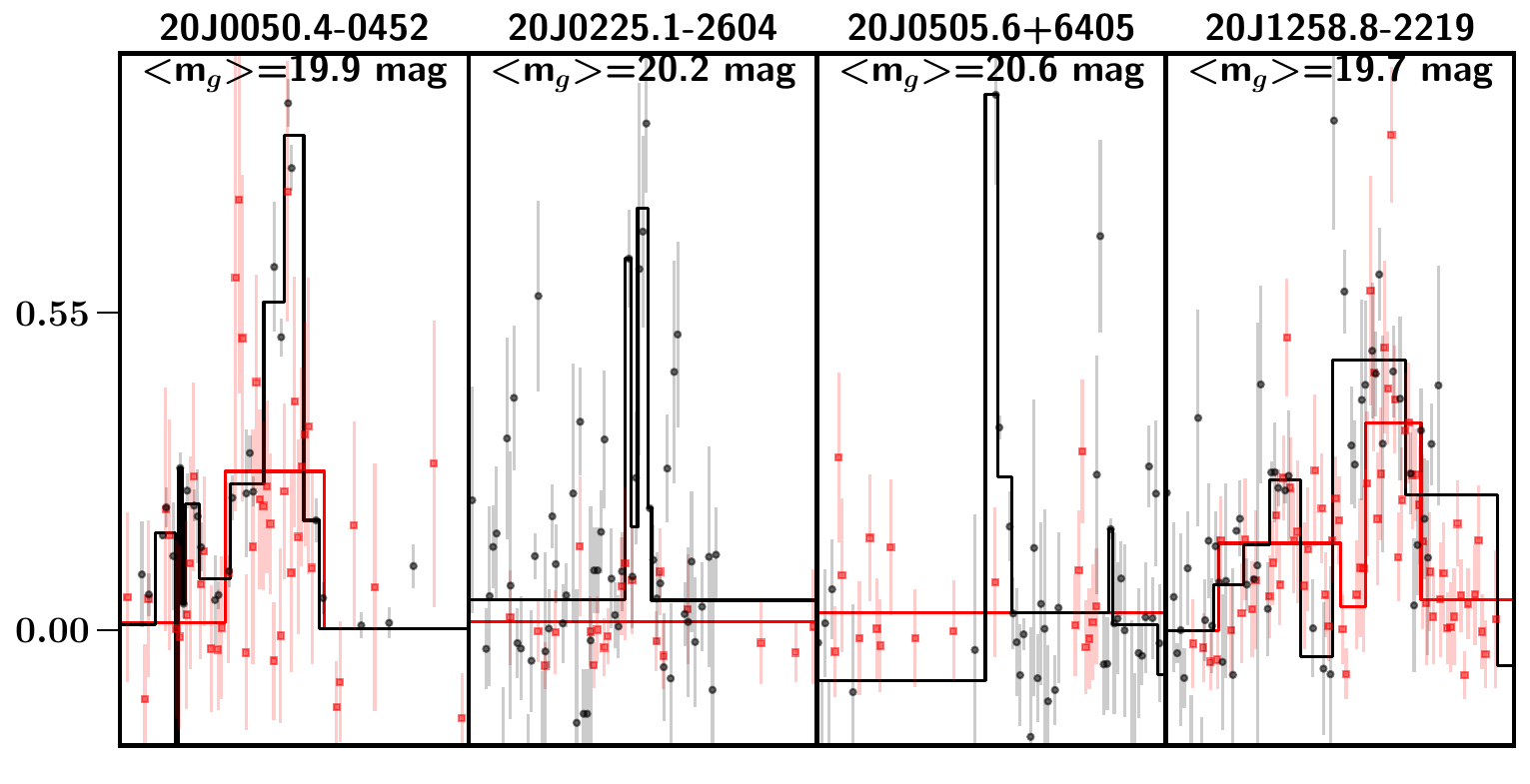}
\caption{Example of optical flares observed in faint optical blazars (m$_{g}$>18.5 mag). In each panel the optical blazar magnitude is shown. Black dots and red squares represent the $V$-band and $\gamma$-ray light curves and the red and black lines are the Bayesian Block Decompositions for each band. For 20J0050.4-0452 and 20J1258.8-2219 we clearly see a $\gamma$-ray counterpart.}
\label{fig:faint_flares}%
\end{figure*}

Finally, we investigate how the statistics are affected by our instrument limiting magnitude. Figure \ref{fig:mag_lim} shows the histogram of a rough estimate of the $g$-band magnitude for blazars with (red) or without optical flares (blue). The $g$-band magnitudes are derived using an average weighted $g$-band flux over all the years (negative and positive values) and a zero point of 16.4. As expected, blazars without optical flares are on average fainter than those with flares, almost one magnitude. Also, as seen in Figure \ref{fig:mag_lim} (right panel), brighter blazars have more detected flares than fainter objects ($\sim$ 6 versus 3 flares). Therefore, as a conservative approach, we select only the optical flares in blazars with a mean $m_{g}<18.5$ mag. This corresponds to our limiting magnitude and the magnitude where the number of flares per object flattens (see Figure \ref{fig:mag_lim}). This is a conservative approach and some of the optical flares detected in fainter blazars are well defined and also correlated with the $\gamma$-ray emission (see Section \ref{txt:tlag}). In Figure \ref{fig:faint_flares}, we present examples of two clear optical flares and two correlated flares in fainter blazars.

After all the cuts, we end up with 1230 optical and 510 $\gamma$-ray flares, so, we detect $\sim$ 42\% more flares in optical than in $\gamma$-ray. This trend has also been seen in \citet{liodakis18}, where they found $\sim$ 3.5 times more optical flares than $\gamma$-ray events. However, in a more recent study, \citet{liodakis19} increased their previous sample (178 versus 145 blazars) and derived a similar number of flares for each band: 4277 for optical and 4384 for $\gamma$-ray. It is important to note that with the BBD method, the number of flares varies with binning, with more flares for smaller temporal bins \citep{liodakis19}. The flare number is also sensitive to the photometric uncertainties, so the flare numbers change for different bin size and methodologies. This could explain our differences with \citet{liodakis19} and the different flare numbers between the optical and $\gamma$-ray bands. For example, the BBD method can explain the larger optical flare rate, as it creates several small false flares due to the smaller photometric errors and higher dispersion of the optical light curves. If we concentrate our study on only the strongest flares and increase our threshold in Equation \ref{eq:flares} from 3 to 10 (i.e., the flare flux must exceed ten times our quiescent level), we only detect 273 optical flares and 161 $\gamma$-ray flares.  

We also investigate the number of flares for each blazar class. In our sample of 421 blazars (with at least one flare), there are 220 FSRQs, 160 BL Lacs, and 41 BCU. We detect 874, 774, and 92 flares for the FSRQ, BL Lac, and BCU classes. Like \citet{liodakis18}, we find that FSRQs tend to exhibit more $\gamma$-ray flares per source than the BL Lacs with 520/650 optical and 354/124 $\gamma$-ray flares for the FSRQ/BL Lac classes. However, the BL Lacs seem to have more optical flares than the FSRQs. Finally, we consider all the FSRQs and BL Lacs of our sample and not only those with at least one flare. We obtain a rate of 1.73/1.77 flares per source (503/437) for FSRQ/BL Lac classes. For optical and $\gamma$-ray, the rates are 1.03/1.48 and 0.70/0.28 flares per source for FSRQ/BL Lac classes, respectively.

\subsection{Time lags}\label{txt:tlag}

Time lags between optical and $\gamma$-ray flares are of the order of only a few days \citep{bonning12,cohen14,liodakis19,bhatta21}, so we look for correlated flares within a 30-day rest-frame window. For each of these 175 objects with flares in both bands, the final rest-frame $\tau_{lag}$ is the average of correlated flare lags, and the error is the quadratic sum of the block widths. In the cases where two flares in one band could be correlated with a simple flare in the other band (i.e., within 30 days), we include both time lags. Our final rest-frame $\tau_{lag}$ and their associated errors are in Table \ref{tab:sample}. Unlike Section \ref{txt:res_flares}, here we do not remove the optical flares seen in faint blazars because we only consider correlated flares. Even in faint blazars, if a $\gamma$-ray emission is seen at roughly the same epoch as the optical flares, it is likely that the optical flare is real. An example of two correlated flares in faint blazars are displayed in Figure \ref{fig:faint_flares}.

\begin{figure}
\centering
\includegraphics[width=0.5\textwidth]{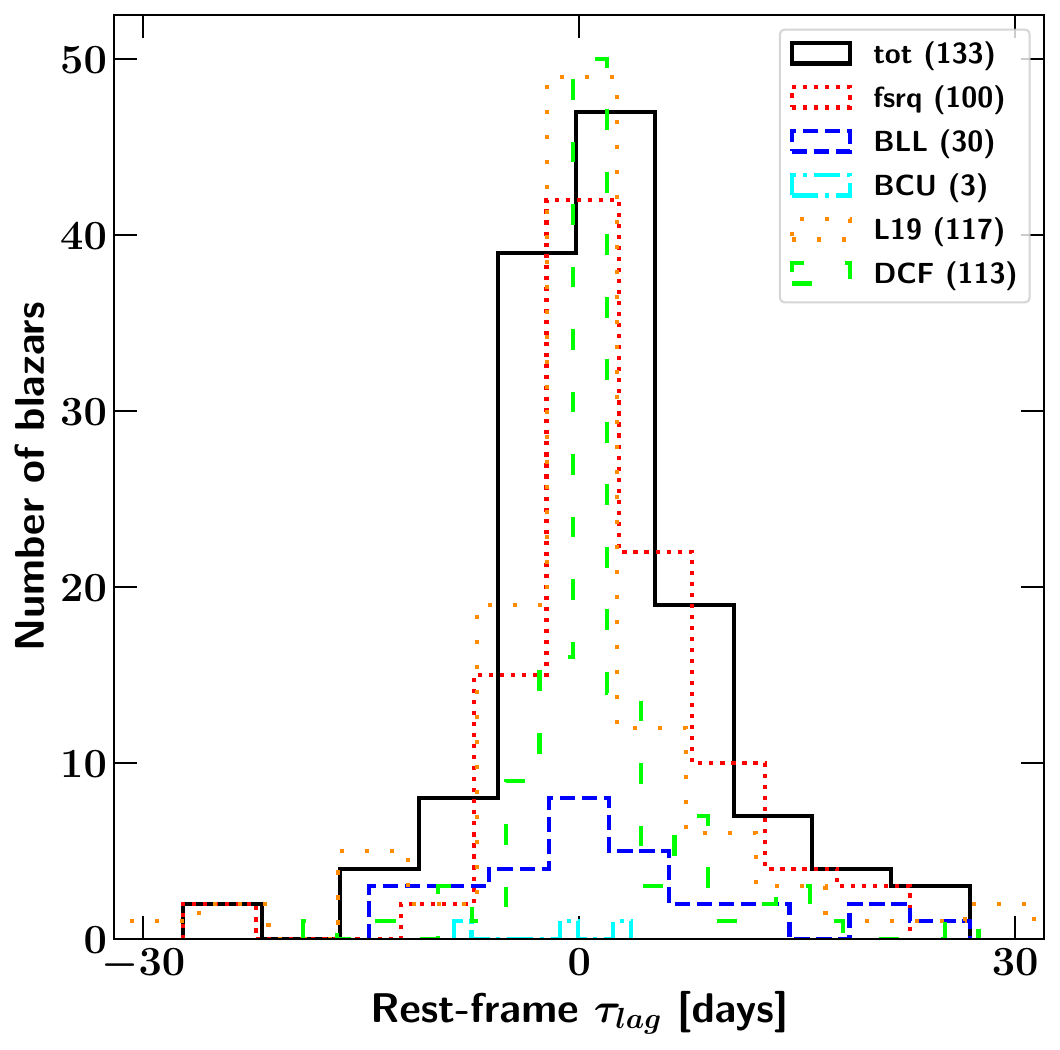}
\caption{Distribution of all rest-frame time lags between the optical and $\gamma$-ray flares (solid black). The distributions for the classes are also shown: FSRQ (dotted red), BL Lac (dashed blue), and BCU (dash-dotted cyan). For comparison, the result obtained by \citet{liodakis19} is shown in loosely dotted orange, and the DCF lag distribution derived from the DCF is the loosely dashed lime. A positive $\tau_{lag}$ corresponds to the $\gamma$-ray emission leading the optical emission.}
\label{fig:tlag}%
\end{figure}

From the 175 objects with flares in both bands, we found 133 blazars with correlated flares and a redshift. Figure \ref{fig:tlag} shows the rest-frame time lag distribution between both bands. We obtain an average rest-frame time lag of 1.1$_{-8.5}^{+7.1}$ days, and the quoted uncertainties (hereafter) represent the 68\% confidence interval obtained from a bootstrapping simulation of 10,000 events. To estimate the uncertainties from the bootstrapping, we select 133 random rest-frame time lags for each event from our distribution and measure the standard deviation. Finally, from those 10,000 measurements, we derive our 68\% confidence interval. Note that, if we simply use the scatter in the 10,000 median lags, the dispersion is only $\sim$ 1 day which it is not representative of our distribution (standard deviation $\sim$ 7--8 days). Among the 133 objects, only three have a rest-frame time lag greater than 20 days and are not consistent with 0 (20J0133.1$-$5201, 20J1006.7$-$2159, 20J1604.6+5714). Our distribution is almost centred on 0 days and it is consistent with previous works \citep{bonning12,cohen14,liodakis19,bhatta21}. For example, \citet{liodakis19} derived a median time lag of $-$0.24 $\pm$ 20.5 days using 117 objects, so, it is consistent with no time lag. Their time lag distribution is similar to ours but has a tail towards longer time lags (see Figure \ref{fig:tlag}). However, the difference may be due to our 30 day search window. If we increase our window to 60 days, the number of objects increases to 143, extending the distribution from $-$55 days to 56 days. However, the median value changes only to 2.9$_{-13.1}^{+15.9}$ days. We also find no correlations between our time lags and the redshift or the synchrotron peak frequency. If we remove the 17 blazars with m$_{g}$>18.5 mag, we do not see significant difference in our average rest-frame time lag (0.2$_{-10.1}^{+5.0}$).

Investigating time lags between flares in different energy bands probes the origin of the seed photon for the leptonic model. For example, using 13 FSRQs and 17 BL Lacs, \citet{cohen14} found that FSRQs tend to have $\gamma$-rays leading the optical by a few days, while for the BL Lacs, they did not find a clear offset. This suggests that the seed photons for FSRQs are from external sources, while it is within the jet for BL Lacs \citep{bottcher13}. However, \citet{liodakis19} did not find such differences using a larger sample (53 FSRQs and 67 BL Lacs).

Figure \ref{fig:tlag} shows the time lag distribution of our 100 FSRQs and 30 BL Lacs with median values of 0.9$_{-8.0}^{+6.3}$ and 1.5$_{-10.9}^{+8.2}$ days for FSRQs and BL Lacs, respectively. A Kolmogorov-Smirnov test also confirms the distribution are consistent with each other. These results support \citet{liodakis19} in finding no evidence for a difference in the seed photon source of the two blazar classes. However, this is only a statistical statement; individual objects could have different high-energy emission mechanisms, as has already been seen for individual blazars (e.g., \citealt{punch92,dermer93,mannheim93,urry97,ghisellini98,mucke01,blazejowski05,aharonian07,albert07,jorstad10,bottcher13,cerruti19}). 

We also separate our sample based on their SED class (LSP, ISP, and HSP). Of 133 objects with $\tau_{lag}$ measurements, we have 118 LSP (96 FSRQs, 19 BL Lacs, 3 BCU), five ISP (1 FSRQ, 4 BL Lacs), and seven HSP (all BL Lacs), and three without an SED class. We derive a median lags of $\tau_{lag}$ of 0.90$_{-8.1}^{+6.2}$ days, 3.6$_{-4.6}^{+1.8}$ days, and 5.6$_{-10.3}^{+15.1}$ days for the LSP, ISP, and HSP classes, respectively. We do not see any significant lags between the optical and $\gamma$-ray flares for the three SED classes. All are consistent with no lag given their uncertainties.  

Finally, as different studies have used the DCF to derive time lags, Figure \ref{fig:tlag} shows our DCF lag distribution for the 113 objects with well measured DCF peaks. The DCF sample is smaller because some sources have correlated flares, but the DCF does not show a well-defined peak (e.g., Figure \ref{fig:dcf}). The DCF lag distribution is also statistically consistent with the lag distribution from the correlated peak analysis (Pearson correlation factor of 0.33, p-value of 0.009) and it has a median rest-frame $\tau_{lag}$ of 0.6$_{-6.2}^{+4.8}$ days.

To summarise our time lag analysis, we find that the majority of blazars, have a strong optical/$\gamma$-ray correlation with timescales on the order of days to tens of days, and are on average consistent with no delay for all blazar classes. Our results are in good agreement with previous studies \citep{bonning12,cohen14,liodakis19,bhatta21} and our work supports the leptonic single-zone models of blazar flares as the driver for non-orphan flares.

\subsection{Orphan flares}\label{txt:res_orph}

Our study of time lags presented in Section \ref{txt:tlag} supports a leptonic single-zone model for the non-orphan flares. However, some blazars show behaviours not expected in such a model. One of the most peculiar is the presence of ``orphan'' flares -- $\gamma$-ray flares with no visible counterpart or optical flares without $\gamma$-ray counterparts \citep{krawczynski04,blazejowski05}. As in Section \ref{txt:res_flares}, we remove all the ``orphan'' flare candidates which occur during optical or $\gamma$-ray observational gaps\footnote{We have only upper limits.} (see Figure \ref{fig:BBD_HOP}). This cut removes 979 ``orphan'' flare candidates: 500 optical (29\%) and 479 $\gamma$-ray (48\%). 

To be sure that the existence of ``orphan'' flares is not due to sensitivity limitations, we compare the optical and $\gamma$-ray flare amplitudes using all the correlated flares from Section \ref{txt:tlag}. Then, we measure the amplitude ratio $A_{opt}/A_{\gamma}$. We do not find a strong correlation between the amplitudes (Pearson factor of 0.18, p-value of 0.002). Thus to estimate the expected $\gamma$-ray flux amplitude for any ``orphan'' optical flare candidate, we assume the median of 1.3$_{-1.2}^{+1.6}$ of the $A_{opt}/A_{\gamma}$ distribution. This value is consistent with \citet{liodakis19}, who also found that optical flares have a larger average amplitude than $\gamma$-ray flares ($A_{opt}/A_{\gamma} \sim 2.7$). 

\begin{figure*}
\centering
\subfloat[optical ``orphan'' flares]{\includegraphics[width=0.45\textwidth]{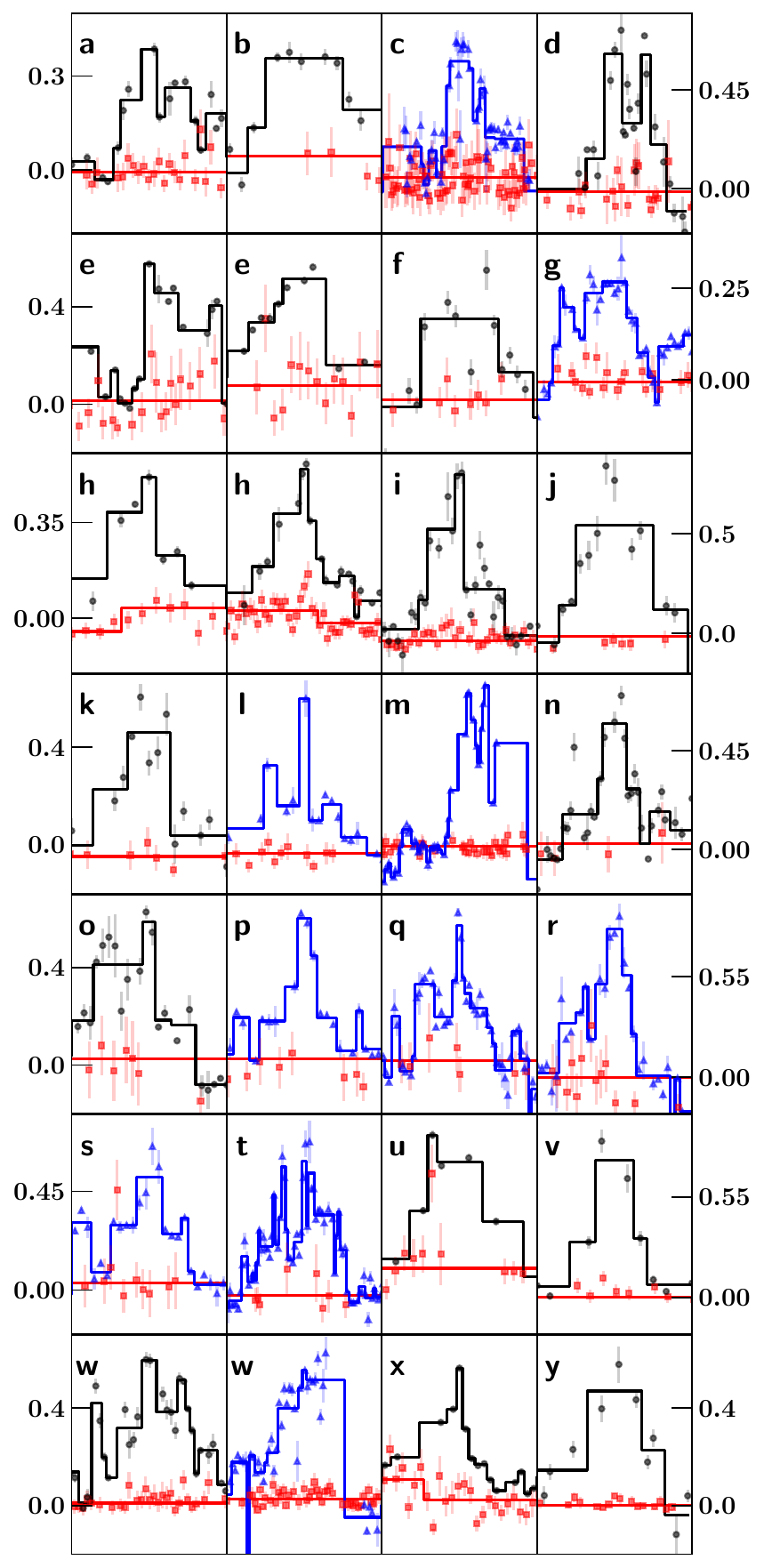}}
\subfloat[$\gamma$-ray ``orphan'' flares]{\includegraphics[width=0.45\textwidth]{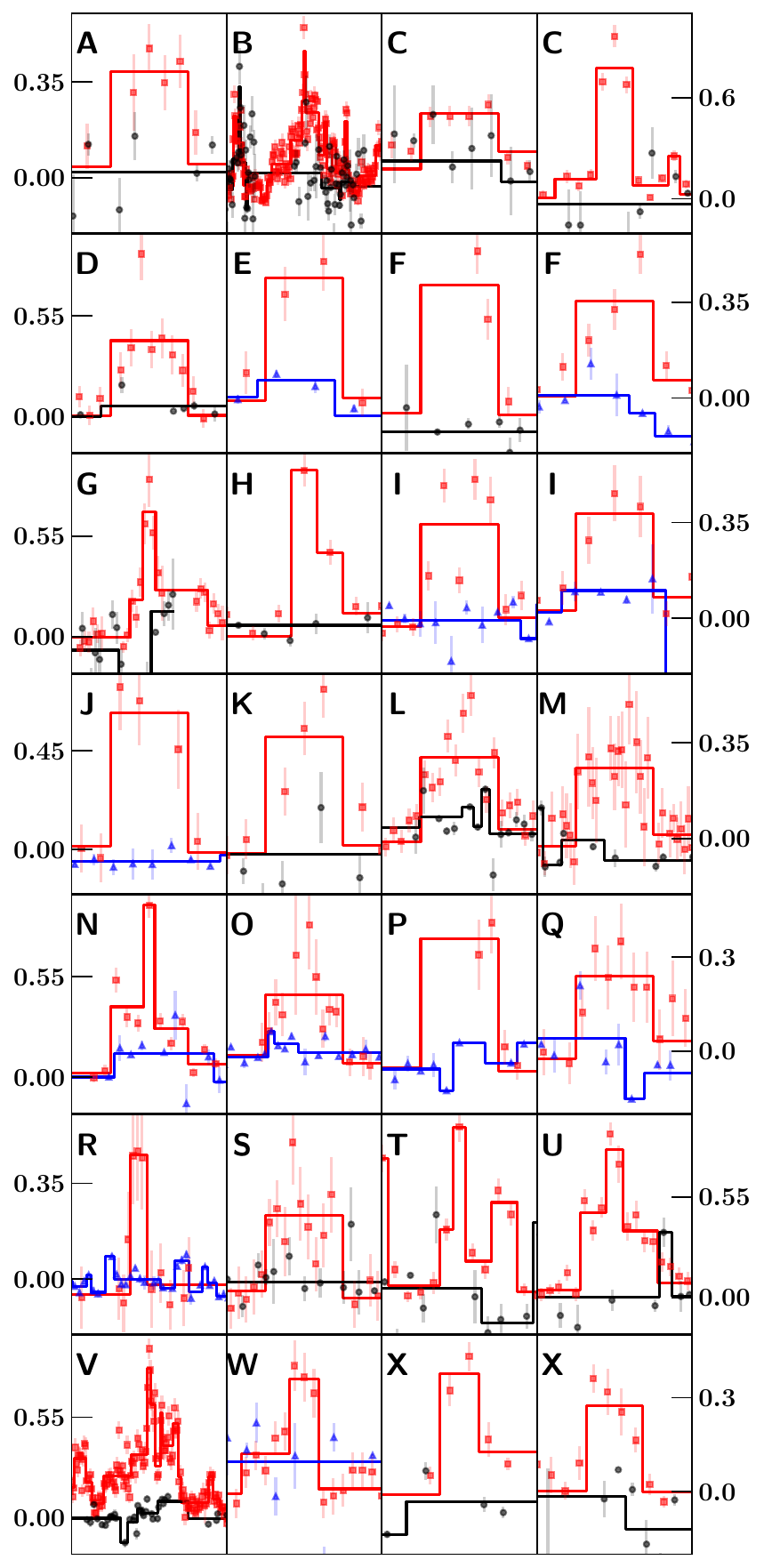}}
\caption{Gold sample of optical (left) and $\gamma$-ray (right) ``orphan'' flares. Black dots, blue triangles, and red squares are optical $V$, optical $g$, and $\gamma$-ray fluxes, respectively. The letter in each panel corresponds to the blazar name in Table \ref{tab:orphan}.}
    \label{fig:orphans}
\end{figure*}

Next, we visually inspect all the ``orphan'' flare candidates and select only the clearest ones, those with at least 2 points, not close to observational gaps, not likely due to photometric noise (see Figure \ref{fig:BBD_HOP}), and seen only in one optical band during $V/g$ bands overlap period. We found a total of $\sim$ 306 ``orphan'' flares: 191 ``orphan'' optical flares and 115 ``orphan'' $\gamma$-ray flares. From those 306 ``orphan'' flares, we also construct a ``gold'' sample of flares for future individual analysis. Figure \ref{fig:orphans} shows those 28 ``orphan'' optical flares and 28 ``orphan'' $\gamma$-ray flares. Our fraction of ``orphan'' $\gamma$-ray flares (115) relative to the total number of $\gamma$-ray flares (507) is consistent with \citet{liodakis19}. We estimate that $\sim$22\% of $\gamma$-ray flares are ``orphan'' events compared to $\sim$20\% in \citet{liodakis19}. This is not surprising as we use the same photometric data for the $\gamma$-ray light curves (from \textit{Fermi}-LAT). However, we find a different fraction percentage of ``orphan'' optical flares. We find that $\sim$13.5\% (191/1414) of optical flares are orphan events as compared to $\sim$54\% in \citet{liodakis19}. A difference in methodology and photometric data could explain the difference. First, we only use the strongest flares, while \citet{liodakis19} considered all local maxima (centre of three Bayesian blocks) as flares, independent of their amplitudes. With a blazar sample roughly three times larger, we find only $\sim$ 2,000 flares while they detected more than 8,600 flares. Second, unlike the $\gamma$-ray data, we are using optical photometry from different telescopes with different limiting magnitudes. We use data from ASAS-SN, which has a limiting magnitude of 18.5 mag \citep{kochanek17b}. Most of the optical light curves used in \citet{liodakis19} are from the KAIT telescope \citep{filippenko01} with a limiting magnitude of 19.5 mag, which allows \citet{liodakis19} to detect fainter optical flares which presumably leads to more ``orphan'' optical flares. \citet{liodakis18} also found that ``orphan'' optical flares in their sample have lower amplitudes and are likely random fluctuations, not real flares. Note also, if we remove all the ``orphan'' optical flares in the blazars with m$_{g}$>18.5 mag, our `orphan'' optical flare fraction decreases to 12\% (from 191 to 153).

 
We also estimate the ``orphan'' flare fraction for each blazar class. We obtained a similar optical ``orphan'' flare number for the two classes  (91/91) but we see a different rates of ``orphan'' $\gamma$-ray flares (85/18) for the FSRQ/B LLac classes. Given the number of FSRQs (220) and BL Lacs (160) in our sample ($+$ 41 BCU), we find more ``orphan'' $\gamma$-ray flares among the FSRQs than among the BL Lacs: 0.38/0.11 ``orphan'' $\gamma$-ray flare per object for FSRQ/BL Lac classes. However, we obtain more ``orphan'' optical flares in BL Lacs than FSRQs: 0.41/0.57 ``orphan'' optical flare per object for FSRQ/BL Lac classes. The presence of ``orphan'' flares argues for multi-zone synchrotron sites or a hadronic origin for some blazar flares. However, as suggested by \citet{liodakis19}, the small orphan $\gamma$-ray flare fraction favours a single-zone leptonic mechanism for the high-energy emission in most blazars.

\subsection{Fractional Variability}

We use the fractional variability to characterise the average variability of the blazars \citep{vaughan03}. It is defined as

\begin{ceqn}
\begin{align}
F_{var}=\sqrt{\frac{S^{2}-\left \langle \sigma _{err}^{2} \right \rangle}{\left \langle F \right \rangle^{2}}} ,
\label{eq:FR}
\end{align}
\end{ceqn}

\noindent where S$^{2}=(N-1)^{-1}\sum_{i=1}^{N}(F_{i}-<F>)^{2} $ is the variance, <F> is the mean flux, and $<\sigma_{err}^{2}>$ is the mean square uncertainty. First, we compare the variability properties using the whole light curves. We measure the variability for 166 objects among the 175 with flares in both bands (for 9 blazars S$^{2}$ is smaller than $<\sigma_{err}^{2}>$). We obtain a mean value and a standard deviation of F$_{var}$=0.60 $\pm$ 0.42 and F$_{var}$= 0.66 $\pm$ 0.34 for the optical and $\gamma$-ray bands, respectively. Even if both bands have a consistent fractional variability, a Kolmogorov-Smirnov test indicates that the fractional variability distributions are not the same ($p=0.0078$). For the different blazar classes, we find F$_{var}$= 0.66 $\pm$ 0.44 and F$_{var}$=0.73 $\pm$ 0.33 for FSRQs and the optical and $\gamma$-ray bands, respectively and F$_{var}$=0.45 $\pm$ 0.27 and F$_{var}$=0.49 $\pm$ 0.25 for BL Lacs. For the $\gamma$-ray band and FSRQs, our fractional variability values are higher but consistent with the value of 0.55 $\pm$ 0.33 found by \citet{rajput20}. We also find that the FSRQ class has a higher fractional variability than the BL Lac class in both the $\gamma$-ray and optical. A Kolmogorov-Smirnov test shows that FSRQ and BL Lac distributions are not from the same distribution at 95\% confidence with statistics of 0.39/0.34 and a p-value of 0.00006/0.0008 (for $\gamma$-ray/optical).

Finally, because F$_{var}$ is sensitive to the photometric uncertainties, we checked our analysis by doubling the uncertainties of both bands. We find different values for F$_{var}$ but the FSRQs still have larger variability in both bands (Opt: FSRQs: 0.63 $\pm$ 0.45, BL Lacs: 0.55 $\pm$ 0.24; $\gamma$-ray FSRQs: 0.79 $\pm$ 0.35 , BL Lacs: 0.65 $\pm$ 0.27).

\section{Conclusions}

Using blazar light curves from the optical All-Sky Automated Survey for Supernovae and the $\gamma$-ray \textit{Fermi}-LAT telescope, we performed the most extensive statistical correlation study between both bands to date for a sample of 1,180 blazars. This is almost an order of magnitude larger than other recent studies \citep{cohen14,liodakis19}. We decompose both light curves into histograms using the Bayesian Block Decomposition method. Then, we select only the prominent flares to study the time delays and the relative rates of ``orphan'' flares. Of the 1,180 blazars, 421 objects have at least one flare and 133 sources have at least one correlated flare within a 30-day window. Our six main conclusions are:

\begin{enumerate}

\item{Based on the 143 sources with correlated flares, the time lag distribution between both bands is consistent with no lags (16th, 50th, and 84th percentiles of $-$8.5, 1.1, and 7.1 days).\\}

\item{We do not find differences in the time lags for different blazar classes. Both distributions are consistent, with median value of 0.9$_{-8.0}^{+6.3}$ and 1.5$_{-10.9}^{+8.2}$ days for our 100 FSRQs and 30 BL Lacs, respectively.\\}

\item{In the 421 blazars with flares, we detect $\sim$ 1400 optical and 500 $\gamma$-ray flares. The exact numbers are sensitive to parameter choices but FSRQs tend to exhibit more $\gamma$-ray flares per source than the BL Lacs while BL Lacs have more optical flares than FSRQs.\\}

\item{We measure a median amplitude ratio between optical and $\gamma$-ray flares of $A_{opt}/A_{\gamma}= 1.3_{-1.2}^{+1.6}$.\\}

\item{We search for good example of ``orphan'' flares and found a total of 306 `orphan'' flares: 191 ``orphan'' optical flares and 115 ``orphan'' $\gamma$-ray flares. These represent $\sim$13\% of optical and $\sim$22\% of $\gamma$-ray flares. We find more ``orphan'' $\gamma$-ray flares in FSRQs than BL Lacs but, more ``orphan'' optical flares in BL Lacs than FSRQs.\\}

\item{We compare the variability properties in the optical and $\gamma$-ray bands. Both bands have a consistent fractional variability amplitude, but FSRQs have higher a fractional variability in both the $\gamma$-ray and optical than BL Lacs.\\}

\end{enumerate}

Our work supports the leptonic single-zone models where the low and high-energy emissions come from the same population of electrons as the dominant driver of blazar flares. However, the presence of ``orphan'' flares in some blazars argues for the occasional presence of a more complex emission mechanism, such as multi-zone synchrotron models or a hadronic origin for some flares.

\section*{Acknowledgements}

We thank Las Cumbres Observatory and its staff for their continued support of ASAS-SN. ASAS-SN is funded in part by the Gordon and Betty Moore Foundation through grants GBMF5490 and GBMF10501 to the Ohio State University, and also funded in part by the Alfred P. Sloan Foundation grant G-2021-14192.

T.d.J  thanks Colby Haggerty for discussions and Jedidah Isler for useful discussions in the early stages of this work. Support for T.d.J has been provided by NSF grants AST-1908952 and AST-1911074. B.J.S. is supported by NSF grants AST-1907570, AST-1908952, AST-1920392, and AST-1911074. C.S.K and K.Z.S are supported by NSF grants AST-1814440 and AST-1908570. A.F acknowledges funding from the German Science Foundation DFG, via the Collaborative Reasearch Center SFB1491 ``Cosmic Interacting Matters - From Source to Signal''. Support for TW-SH was provided by NASA through the NASA Hubble Fellowship grant HST-HF2-51458.001-A awarded by the Space Telescope Science Institute, which is operated by the Association of Universities for Research in Astronomy, Inc., for NASA, under contract NAS5-265. J .F.B. is supported by National Science Foundation grant No.\ PHY-2012955. 

This work is based on observations made by ASAS-SN. We wish to extend our special thanks to those of Hawaiian ancestry on whose sacred mountains of Maunakea and Haleakal\=a, we are privileged to be guests. Without their generous hospitality, the observations presented herein would not have been possible.\\

\textit{Facilities:} \textit{Fermi}-LAT telescope, Haleakala Observatories (USA), Cerro Tololo International Observatory (Chile), McDonald Observatory (USA), South African Astrophysical Observatory (South Africa).\\

\textit{Software:} astropy \citep{astropy}, Matplotlib \citep{matplotlib}, Numpy \citep{numpy}, Scipy \citep{scipy}, Z-transformed discrete correlation function \citep{alexander97,alexander13}

\section*{Data Availability Statements}
The optical and $\gamma$-ray light curves have already been made available to the community. $\gamma$-ray light curves can be downloaded from \url{https://fermi.gsfc.nasa.gov/ssc/data/access/lat/LightCurveRepository/source.html?} while optical light-curves from ASAS-SN can be obtained via their ASAS-SN sky patrol webpage (\url{https://asas-sn.osu.edu/}). All the datasets used in this work will be also available on the author's Github (\url{https://github.com/tdejaeger}).




\appendix

\section{Properties of Blazars used in our sample}\label{AppendixA}
In this appendix, we list relevant information from \citet{abdollahi22} and \citet{ballet20} of our blazar sample together with the results obtained in Section \ref{txt:RESULTS}.

\onecolumn
\scriptsize
\begin{longtable}{lcccccccc}
\caption{Blazar properties.}\\
\hline
\hline 
\multicolumn{1}{l}{4FGL Name} & \multicolumn{1}{c}{RA} & \multicolumn{1}{c}{Dec} & \multicolumn{1}{c}{Association}& \multicolumn{1}{c}{Class} & \multicolumn{1}{c}{SED class} & \multicolumn{1}{c}{redshift} & \multicolumn{1}{c}{Flares} & \multicolumn{1}{c}{Lags}\\
\multicolumn{1}{l}{} & \multicolumn{1}{c}{deg} & \multicolumn{1}{c}{deg} & \multicolumn{1}{c}{} & \multicolumn{1}{c}{} & \multicolumn{1}{c}{} & \multicolumn{1}{c}{} & \multicolumn{1}{c}{} & \multicolumn{1}{c}{days}\\
\hline
\endfirsthead
\hline
\hline 
\multicolumn{1}{l}{4FGL Name} & \multicolumn{1}{c}{RA} & \multicolumn{1}{c}{Dec} & \multicolumn{1}{c}{Association}& \multicolumn{1}{c}{Class} & \multicolumn{1}{c}{SED class} & \multicolumn{1}{c}{redshift} & \multicolumn{1}{c}{Flares} & \multicolumn{1}{c}{Lags}\\
\multicolumn{1}{l}{} & \multicolumn{1}{c}{deg} & \multicolumn{1}{c}{deg} & \multicolumn{1}{c}{} & \multicolumn{1}{c}{} & \multicolumn{1}{c}{} & \multicolumn{1}{c}{} & \multicolumn{1}{c}{} & \multicolumn{1}{c}{days}\\
\hline
\endhead
 J0001.2-0747 &    0.315100 &  -7.797100 &                PMN J0001-0746 &   BL Lac &       LSP & $\cdots$ &    opt &              $\cdots$ \\
 J0001.5+2113 &    0.381500 &  21.218300 &                  TXS 2358+209 &  FSRQ &       ISP &    1.106 &   both &      0.25 (4.64) \\
 J0003.3-1928 &    0.846500 & -19.467600 &                  PKS 0000-197 &   BCU &       LSP & $\cdots$ &   none &              $\cdots$ \\
 J0004.3+4614 &    1.075700 &  46.242699 &              MG4 J000421+4615 &  FSRQ &       LSP &    1.810 &   none &              $\cdots$ \\
 J0004.4-4737 &    1.109100 & -47.623299 &                  PKS 0002-478 &  FSRQ &       LSP &    0.880 &   none &              $\cdots$ \\
 J0005.9+3824 &    1.498600 &  38.401001 &                    S4 0003+38 &  FSRQ &       LSP &    0.229 &   none &              $\cdots$ \\
 J0007.7+4008 &    1.927500 &  40.133999 &           NVSS J000741+400830 &   BCU & $\cdots$ & $\cdots$ &   none &              $\cdots$ \\
 J0009.3+5030 &    2.346600 &  50.511501 &           NVSS J000922+503028 &   BL Lac &       HSP & $\cdots$ &   none &              $\cdots$ \\
\end{longtable}
\vspace{-0.60cm}
\noindent
\hspace{-0.90cm}
\begin{minipage}[b]{19.5cm}
\scriptsize
Notes: The first column gives the 4FGL name, followed by its coordinates in columns 2 and 3. In Column 4, 5, 6, we list the associated source to the $\gamma$-ray emissions, the sources class (BL Lac: BL Lac type, FSRQ: flat-spectrum radio quasars, BUC: blazar candidates of uncertain type), and the SED class (LSP: low-synchrotron peaked blazars, ISP: intermediate synchrotron peaked blazars, and HSP: high-synchrotron peaked blazars). In Column 7, the redshift is given. In Columns 8, the presence of significant flares in the optical and $\gamma$-rays light curves is display (both, opt, gam, none). Finally, in Column 9, we display the rest-frame time lag in days between optical and $\gamma$-rays flares, where a positive $\tau_{lag}$ corresponds to the $\gamma$-ray emission leading to the optical emission.\\
\end{minipage}

\section{``Orphan'' flares}\label{AppendixB}
\normalsize

In this appendix, we list the characteristic of our ``orphan'' flare golden sample.

\onecolumn
\begin{longtable}{lcccccccc}
\caption{``Orphan'' flare golden sample properties.}\\
\hline
\hline 
\multicolumn{1}{l}{4FGL Name} & \multicolumn{1}{c}{Band} & \multicolumn{1}{c}{Peak} & \multicolumn{1}{c}{width}& \multicolumn{1}{c}{label Fig. \ref{fig:orphans}}\\
\multicolumn{1}{l}{} & \multicolumn{1}{c}{} & \multicolumn{1}{c}{MJD} & \multicolumn{1}{c}{days} & \multicolumn{1}{c}{}\\
\hline
\endfirsthead
\hline
\hline 
20J0011.4+0057	&gam	&57564&	15&	A\\
20J0038.2$-$2459	&opt	&57930&	45&	a\\
20J0050.7$-$0929	&opt	&56477&	19&	b\\
20J0102.8+5824	&opt	&59124&	171&	c\\
20J0108.6+0134	&gam	&57581&	258&	B\\
20J0118.9$-$2141	&opt	&58349&	109&	d\\
20J0133.1$-$5201	&gam	&57764&	12&	C\\
20J0133.1$-$5201	&gam	&58077&	19&	C\\
20J0211.2+1051	&opt	&56882&	48&	e\\
20J0211.2+1051	&opt	&57050&	27&	e\\
20J0217.8+0144	&opt	&57758&	30&	f\\
20J0229.5$-$3644	&gam	&57212&	22&	D\\
20J0303.4$-$2407	&opt	&58701&	66&	g\\
20J0325.7+2225	&gam	&58460&	6&	E\\
20J0405.6$-$1308	&gam	&57574&	22&	F\\
20J0405.6$-$1308	&gam	&59571&	9&	F\\
20J0440.3$-$4333	&gam	&58365&	54&	G\\
20J0457.0$-$2324	&opt	&57037&	16&	h\\
20J0457.0$-$2324	&opt	&57252&	60&	h\\
20J0501.2$-$0158	&gam	&57258&	9&	H\\
20J0538.8$-$4405	&opt	&57744&	51&	i\\
20J0539.9$-$2839	&gam	&59208&	15&	I\\
20J0539.9$-$2839	&gam	&59319&	9&	I\\
20J0719.3+3307	&opt	&57324&	27&	j\\
20J0742.6+5443	&opt	&57106&	27&	k\\
20J0839.8+0105	&opt	&58448&	36&	l\\
20J0842.3$-$6053	&gam	&59600&	12&	J\\
20J0854.8+2006	&opt	&58947&	142&	m\\
20J0910.8+3859	&opt	&57090&	69&	n\\
20J0912.2+4127	&gam	&57318&	12&	K\\
20J0921.6+6216	&gam	&57752&	30&	L\\
20J0942.3$-$0800	&opt	&57722&	37&	o\\
20J1058.4+0133	&gam	&56700&	63&	M\\
20J1153.3$-$1104	&gam	&58467&	21&	N\\
20J1159.2$-$2227	&gam	&59332&	34&	O\\
20J1218.5$-$0119	&opt	&58872&	36&	p\\
20J1238.3$-$1959	&opt	&58545&	84&	q\\
20J1303.0+2434	&opt	&59366&	48&	r\\
20J1308.5+3547	&gam	&58970&	18&	P\\
20J1312.8$-$0425	&gam	&58658&	18&	Q\\
20J1333.7+5056	&gam	&59360&	50&	R\\
20J1423.5$-$7829	&opt	&59271&	36&	s\\
20J1427.6$-$3305	&opt	&58591&	111&	t\\
20J1438.9+3710	&gam	&57863&	30&	S\\
20J1506.1+3731	&gam	&57060&	18&	T\\
20J1517.7$-$2422	&opt	&58197&	25&	u\\
20J1532.7$-$1319	&gam	&56771&	27&	U\\
20J1635.2+3808	&gam	&56499&	201&	V\\
20J1700.0+6830	&opt	&56466&	18&	v\\
20J1740.5+5211	&gam	&59271&	24&	W\\
20J1748.6+7005	&opt	&56864&	55&	w\\
20J1748.6+7005	&opt	&59514&	90&	w\\
20J2158.8$-$3013	&opt	&57604&	37&	x\\
20J2236.5$-$1433	&opt	&58413&	40&	y\\
20J2250.0$-$1250	&gam	&57674&	12&	X\\
20J2250.0$-$1250	&gam	&58340&	16&	X\\
\hline
\label{tab:orphan}
\end{longtable}


\bsp	
\label{lastpage}
\end{document}